\documentclass[pre,twocolumn,superscriptaddress]{revtex4}  

\usepackage[paperwidth=210mm,paperheight=297mm,centering,hmargin=2cm,vmargin=2.5cm]{geometry}

\usepackage[pdftex]{graphicx}
\usepackage{amsmath,amsfonts,amssymb}
\usepackage{morefloats}
\usepackage{xcolor} 
\usepackage{tabularx}

\usepackage{newfloat,algcompatible}
\usepackage{etoolbox}
\DeclareFloatingEnvironment[
    fileext=loa,
    listname=List of Algorithms,
    name= Algorithm,
    placement=tbhp,
]{algorithm}

\makeatletter
\renewcommand{\fnum@algorithm}{\small\textbf{\algorithmname~\thealgorithm}}
\makeatother

\definecolor{gris25}{gray}{0.5}
\definecolor{gris5}{gray}{0.7}
\definecolor{gris75}{gray}{0.9}
\definecolor{green2}{rgb}{0,0.8,0}
\definecolor{bluemoi}{rgb}{0.25,0.50 ,0.75} 

\usepackage{hyperref}
\hypersetup{
  bookmarksopen=false,
  pdftitle=" ",
  pdfauthor=" ", 
  pdfsubject=" ", 
  pdftoolbar=true, 
  pdfmenubar=true, 
  pdfhighlight=/O, 
  colorlinks=true, 
  pdfpagemode=UseNone, 
  pdfpagelayout=SinglePage, 
  pdffitwindow=true, 
  linkcolor=bluemoi, 
  citecolor=bluemoi, 
  urlcolor=bluemoi 
}

\makeatletter
\renewcommand{\figurename}{Figure}
\renewcommand{\fnum@figure}{\small\textbf{\figurename~\thefigure}}
\renewcommand{\thefigure}{\arabic{figure}}
\makeatother

\makeatletter
\renewcommand{\tablename}{Table}
\renewcommand{\fnum@table}{\small\textbf{\tablename~\thetable}}
\renewcommand{\thetable}{\arabic{table}}
\makeatother

\begin{document}

\title{Generating a synthetic population of individuals in households: Sample-free vs sample-based methods}

\author{Maxime Lenormand}\affiliation{IRSTEA, LISC, 24 avenue des Landais, 63172 AUBIERE, France}
\author{Guillaume Deffuant}\affiliation{IRSTEA, LISC, 24 avenue des Landais, 63172 AUBIERE, France}

\begin{abstract} 
We compare a sample-free method proposed by \cite{Gargiulo2010} and a sample-based method proposed by \cite{Ye2009} for generating a synthetic population, organized in households, from various statistics. We generate a reference population for a French region including 1310 municipalities and measure how both methods approximate it from a set of statistics derived from this reference population. We also perform a sensitivity analysis. The sample-free method better fits the reference distributions of both individuals and households. It is also less data demanding but it requires more pre-processing. The quality of the results for the sample-based method is highly dependent on the quality of the initial sample.  
\end{abstract}

\maketitle

\section*{INTRODUCTION}

For two decades, the number of micro-simulation models, simulating the evolution of large populations with an explicit representation of each individual, has been constantly increasing with the computing capabilities and the availability of longitudinal data. When implementing such an approach, the first problem is initialising properly a large number of individuals with the adequate attributes. Indeed, in most of the cases, for privacy reasons, exhaustive individual data are excluded from the public domain. Aggregated data at various levels (municipality, county,...), guaranteeing this privacy, are hence only available in general. Sometimes, individual data are available on a sample of the population, these data being chosen also for guaranteeing the privacy (for instance omitting the individual's location of residence). This paper focuses on the problem of generating a virtual population with the best use of these data, especially when the goal is generating both individuals and their organisation in households.

Two main methods, both requiring a sample of the population, aim at tackling this problem: 
\begin{itemize}
	\item The synthetic reconstruction methods (SR)  \cite{Wilson1976}. These methods generally use the Iterative Proportional Fitting \cite{Deming1940} and a sample of the target population to obtain the joint-distributions of interest \cite{Beckman1996, Huang2002, Guo2007, Arentze2007, Ye2009}. Many of the SR methods match the observed and simulated households joint-distribution or individual joint-distribution but not simultaneously. To circumvent these  limitations \cite{Guo2007, Arentze2007, Ye2009} proposed different techniques to match both household and individual attributes. Here, we focus on the Iterative Proportional Updating developed by \cite{Ye2009}. 
	\item The combinatorial optimization (CO). These methods create a synthetic population by zone using marginals of the attributes of interest and a sub-set of a sample of the target population for each zone (for a complete description see \cite{Voas2000, Huang2002}). 
\end{itemize}

Recently, sample-free SR methods appeared \cite{Gargiulo2010, Barthelemy2012}. The  sample-free SR methods build households by picking up individuals in a set comprising initially the whole population and progressively shrinking.  In \cite{Barthelemy2012}, if there is no appropriate individual in the current set, the individual is picked up in the already generated households, whereas in \cite{Gargiulo2010}, the individuals are picked up in the set only. Both approaches are illustrated on real life examples, \cite{Barthelemy2012} generated a synthetic population of Belgium at the municipality level and \cite{Gargiulo2010} generated the population of two municipalities in Auvergne region (France). These methods can be used in the usual situations where no sample is available and one must only use distributions of attributes (of individuals and households). Hence, they overcome a strong limit of the previous methods. It is therefore important to assess if this larger scope of the sample-free method implies a loss of accuracy compared with the sample-based method.

The aim of this paper is contributing to this assessment. With this aim, we compare the sample-based IPU method proposed by \cite{Ye2009} with the sample-free approach proposed by \cite{Gargiulo2010} on an example.  

In order to compare the methods, the ideal case would be to have a population with complete data available about individuals and households. It would allow us to measure precisely the accuracy of each method, in different conditions. Unfortunately, we do not have such data. In order to put ourselves in a similar situation, we generate a virtual population and then use it as a reference to compare the selected methods as in \cite{Barthelemy2012}. All the algorithms presented in this paper are implemented in JAVA on a desktop machine (PC Intel 2.83 GHz).

In the first section we formally present the two methods. In the second section we present the comparison results. Finally, we discuss our results. 

\section*{Details of the chosen methods}

\subsection*{Sample-free method}

We consider a set of $n$ individuals $X$ to dispatch in a set of $m$ households $Y$ in order to obtain a set of filled households $P$. Each individual $x$ is characterised by a type $t_x$ from a set of $q$ differents individual types $T$ (attributes of the individual). Each household $y$ is characterized by a type $u_y$ from a set of $p$ different household types $U$ (attributes of the household). We define $n_T=\{n_{t_k}\}_{1\leq k \leq q}$ as the number of individuals of each type and $n_U=\{n_{u_l}\}_{1\leq l \leq p}$ as the number of households of each type. Each household $y$ of a given type $u_y$ has a probability to be filled by a subset of individuals $L$, then the content of the household equals $L$, which is denoted $c(y)=L$. We use this probability to iteratively fill the households with the individuals of $X$.  

\begin{equation}
	\mathbb{P}(c(y)=L|u_y)
\label{eq1} 
\end{equation}

The iterative algorithm used to dispach the individuals into the households according to the Equation \ref{eq1} is described in Algorithm \ref{PopGenFlo1}. The algorithm starts with the list of individuals $X$ and of the households $Y$, defined by their types. Then it iteratively picks at random a household, and from its type and Equation \ref{eq1}, derives a list of individual types. If this list of individual types is available in the current list of individuals $X$, then this filled household is added to the result, and the current lists of individuals and households are updated. This operation is repeated until one of the lists $X$ or $Y$ is void, or a limit number of iterations is reached.  

\begin{algorithm}
  {\hrulefill}
	\vspace*{-0.3cm}
	\caption{The general iterative algorithm}
	{\vspace*{-0.25cm}\hrulefill}
	\label{PopGenFlo1}
	\begin{algorithmic}
	  \REQUIRE $X$ and $Y$
	  \ENSURE $P$
	  \STATE Set $P=\varnothing$
		\WHILE{$Y \neq \varnothing$}
		  \STATE Pick at random $y$ from $Y$
			\STATE Pick at random $L$ with a probability defined in 
			\STATE Equation \ref{eq1}
		  \IF{$L \subset X$}	
		  	\STATE $P \leftarrow P \cup L$ 
				\STATE $Y \leftarrow Y\backslash \{y\}$
				\STATE $X \leftarrow X\backslash L$	  	
		  \ENDIF
		\ENDWHILE
	\end{algorithmic}
	{\vspace*{-0.2cm}\hrulefill}
\end{algorithm}

In the case of the generation of a synthetic population, we can replace the selection of the list $L$ by the selection of the individuals one at a time by order of importance in the household. In this case Equation \ref{eq2} replaces Equation \ref{eq1}. 

\begin{equation}
	\begin{array}{l}
		\mathbb{P}(x_1 \in y|u_{y}) \times\\
	  \mathbb{P}(x_2 \in y|u_{y},x_1 \in y)\times\\
	  \mathbb{P}(x_3 \in y|u_{y},x_1 \in y,x_2 \in y)\times\\
	  ...\\
	\end{array}
\label{eq2} 
\end{equation}

The iterative approach algorithm associated with this probability is described in Algorithm \ref{PopGenFlo2}. The principle is the same as previously, it is simply quicker. Instead of generating the whole list of individuals in the household before checking it, one generates this list one by one, and as soon as one of its members cannot be found in $X$, the iteration stops, and one tries another household.

\begin{center}
\begin{algorithm}
  {\hrulefill}
	\vspace*{-0.3cm}
	\caption{The iterative algorithm}
	{\vspace*{-0.25cm}\hrulefill}
	\label{PopGenFlo2}
	\begin{algorithmic}
	  \REQUIRE $X$ and $Y$
	  \ENSURE $P$
	  \STATE Set $P=\varnothing$
		\WHILE{$Y \neq \varnothing$}
		  \STATE Pick at random $y$ from $Y$
			\STATE Pick at random $x_1$ with a probability 
			\STATE $\mathbb{P}(x_1 \in y|u_{y})$
			\STATE Pick at random $x_2$ with a probability 
			\STATE $\mathbb{P}(x_2 \in y|u_{y},x_1 \in y)$
			\STATE Pick at random $x_3$ with a probability 
			\STATE $\mathbb{P}(x_3 \in y|u_{y},x_1 \in y,x_2 \in y)$
			\STATE ...
		  \IF{$\{x_1,x_2,x_3,...\} \subset X$}
		  	\STATE $P \leftarrow P \cup \{x_1,x_2,x_3,...\}$ 
				\STATE $Y \leftarrow Y\backslash \{y\}$
				\STATE $X \leftarrow X\backslash \{x_1,x_2,x_3,...\}$	 		  	
		  \ENDIF
		\ENDWHILE
	\end{algorithmic}
	{\vspace*{-0.2cm}\hrulefill}
\end{algorithm}
\end{center}

In practice this stochastic approach is data driven. Indeed, the types $T$ and $U$ are defined in accordance with the data available and the complexity to extract the distribution of the Equation \ref{eq2} increases with $n_T$ and $n_U$. The distributions defined in Equation \ref{eq2} are called distributions for affecting individual into household. In concrete applications, it occurs that one needs to estimate $n_T$, $n_U$ and the distributions of probabilities presented in Equation \ref{eq2}. This estimation implies that the Algorithm \ref{PopGenFlo2} can not converge in a reasonable time because of the stopping criterion ($Y \neq \varnothing$). This stopping criterion is equivalent to an infinite number of "filling" trials by households. In this case, we can replace the stopping criterion by a maximal number of iterations by households and then put the remaining individuals in the remaining households using relieved distributions for affecting individual into household.

In a perfect case where all the data are available and the time infinite, the algorithm would find a perfect solution. When the data are partial and the time constrained, it is interesting to assess how this method manages to make the best use of the available data.

\begin{table*}
	\caption{IPU Table. The light grey table represents the frequency matrix $D$ showing the household type $U$ and the frequency of different individual types $T$ within each filled households for the sample $P_s$. The dimension of $D$ is $|P_s|\times(p+q)$, where $|P_s|$ is the cardinal number of the sample $P_s$, $q$ the number of individual types and $p$ the number of household types. An element $d_{ij}$ of $D$ represents the contribution of filled household $i$ to the frequency of individual/household type $j$.}
	\label{D}		
	\begin{center}
    \begin{tabular}{c}
			\includegraphics[width=\linewidth]{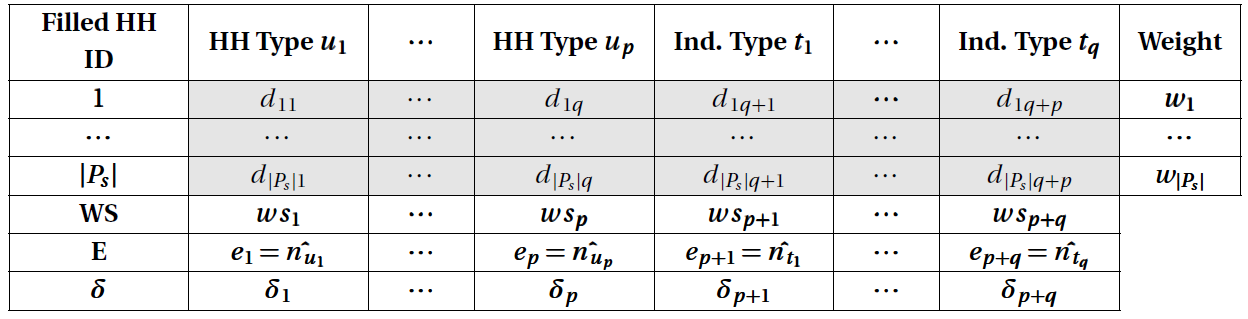}
		\end{tabular}
	\end{center}
\end{table*}

\subsection*{The sample-based approach (General Iterative Proportional Updating)}

This approach, proposed by \cite{Ye2009}, starts with a sample $P_s$ of $P$ and the purpose is to define a weight $w_i$ associated with each individual and each househld of the sample in order to match the total number of each type of individuals in $X$ and households in $Y$ to reconstruct $P$. The method used to reach this objective is the Iterative Proportional Updating (IPU). The algorithm proposed in \cite{Ye2009} is described in Algorithm \ref{IPU}. In this algorithm, for each type of households or individuals $j$ the purpose is to match the weighted sum $sw_j$ with the estimated constraints $e_j$ with an adjustement of the weights. $e_j$ is an estimation of the total number of households or individuals $j$ in $P$. This estimation is done separately for each individual and household type using a standard IPF procedure with marginal variables. When the match between the weighted 

\begin{center}
\begin{algorithm}
  {\hrulefill}
	\vspace*{-0.35cm}
	\caption{Iterative Proportional Updating algorithm}
	{\vspace*{-0.25cm}\hrulefill}
	\label{IPU}
	\begin{algorithmic}
	  \REQUIRE $P_s$, $\epsilon$ 
	  \ENSURE $P$
	  \STATE Set $P=\varnothing$
	  \STATE Generate $D \in \mathrm{M}_{|P_s|\times (p+q)}(\mathbb{R})$ described by the light grey table in Table \ref{D}
	  \STATE Estimate $n_T$ and $n_U$ using the standard IPF procedure and store the resulting estimate into a vector $E=(e_j)_{1\leq j \leq p+q}$ as in Table \ref{D}
	  \FOR{$i=1$ to $|P_s|$}
	  	\STATE Set $w_i=1$
	  \ENDFOR
	  \FOR{$j=1$ to $p+q$}
	  	\STATE Compute $sw_j=\sum_{i=1}^{|P_s|} d_{ij}w_i$
	  	\STATE Compute $\delta_j= \frac{|sw_j-e_j|}{e_j}$ 
	  \ENDFOR
	  \STATE Compute $\delta=\frac{1}{p+q}\sum_{j=1}^{p+q} \delta_j$
	  \STATE Set $\delta_{\mbox{min}}=\delta$
	  \STATE Set $\Delta=\epsilon+1$
	  \WHILE{$\Delta>\epsilon$}
	    \STATE Set $\delta_{\mbox{prev}}=\delta$
	  	\FOR{$j=1$ to $p+q$}
	  	  \FOR{$i=1$ to $|P_s|$}
	  			\IF{$d_{ij}\neq 0$}
	  				\STATE $w_i=\frac{e_j}{sw_j}w_i$
	  			\ENDIF
	  		\ENDFOR	  	  
	  		\STATE Compute $sw_j=\sum_{i=1}^{|P_s|} d_{ij}w_i$	  		
	  	\ENDFOR
	  	\STATE Compute $\delta=\frac{1}{p+q}\sum_{j=1}^{p+q} \delta_j$
	    \IF{$\delta < \delta_{\mbox{min}}$}
	  			\STATE Set $W_{\mbox{opt}}=(w_i)_{1\leq i \leq |P_s|}$ 
	  			\STATE $\delta=\delta_{\mbox{min}}$
		 	\ENDIF	
	  	\STATE $\Delta=|\delta-\delta_{\mbox{prev}}|$
	  \ENDWHILE 	  
		\end{algorithmic}
		{\vspace*{-0.2cm}\hrulefill}
\end{algorithm}
\end{center}

\noindent sample and the constraint becomes stable, the algorithm stops. The procedure then generates a synthetic population by drawing at random the filled households of $P_s$ with probabilities corresponding to the weights. This generation is repeated several times and one chooses the result with the best fit with the observed data. 

\section*{Generating a synthetic population of reference for the comparison}

Because we cannot access any population with complete data available about individuals and households, we generate a virtual population and then use it as a reference to compare the selected methods as in \cite{Barthelemy2012}. 

We start with statistics about the population of Auvergne (French region) in 1990 using the sample-free approach presented above. The Auvergne region is composed of 1310 municipalities, 1,321,719 inhabitants gathered in 515,736 households. Table \ref{summary} presents summary statistics on the Auvergne municipalities. 

\begin{table}[!ht]
	\caption{Summary statistics on the Auvergne municipalities}
	\label{summary}
	\begin{center}
		\begin{tabular}{|>{\centering}m{2cm}|>{\centering}m{1.5cm}|>{\centering}m{1.5cm}|m{1.5cm}<{\centering}|}
		\hline
		\textbf{Statistics} & Min  & Max & Average\\
			\hline
		  Households & 8 & 63,226 &  408.2 \\
		  \hline
		  Individuals & 26 & 136,180 & 1,011.7 \\
		  \hline	 
		\end{tabular}
	\end{center}
\end{table} 

\begin{table*}
	\caption{Data description}
	\label{Controls}
	\begin{center}
		\begin{tabular}{|>{\centering}m{1cm}|>{}m{9cm}|m{3cm}<{}|}
		\hline
		\textbf{ID} & \textbf{Description}  & \textbf{Level}\\
			\hline
		  1  & Number of individuals grouped by ages & Municipality (LAU2)\\
		  \hline
		  2  & Distribution of individual by activity status according to the age & Municipality (LAU2)\\
		  \hline
		  3  &	Joint-distribution of household by type and size & Municipality (LAU2)\\	
		  \hline	 
			4  & Probability to be the head of household according to the age and the type of household & Municipality (LAU2)\\
			\hline
			5  & Probability of having a couple according to the difference of age between the partners (from"-16years" to "21years") & National level\\
			\hline
			6  & Probability to be a child (child=live with parent) of household according to the age and the type of household & Municipality (LAU2)\\
			\hline
		\end{tabular}
	\end{center}
\end{table*}

\subsection*{Generation of the individuals}

For each municipality of the Auvergne region we generate a set $X$ of individuals with a stochastic procedure. For each individual of the age pyramid (distribution 1 in Table \ref{Controls}), we randomly choose an age in the bin and then we draw randomly an activity status according to the distribution 2 in Table \ref{Controls}. 

\subsection*{Generation of the households}

For each municipality of the Auvergne region we generate a set $Y$ of households according to the total number of individual $n=|X|$ with a stochastic procedure. We draw at random households according to the distribution 3 in Table \ref{Controls} while the sum of the capacities is below $n$ and then we determine the last household to have $n$ equal to the sum of the size of the households.

\subsection*{Distributions for affecting individual into household}

\noindent{\bf{Single}}

	\begin{itemize}
		\item The age of the individual 1 is determined using the distribution 4 in Table \ref{Controls}.
	\end{itemize}
	
\noindent{\bf{Monoparental}}
	\begin{itemize}
		\item The age of the individual 1 is determined using the distribution 4 in Table \ref{Controls}.
		\item The ages of the children are determined according to the age of individual 1 (An individual can do a child after 15 and before 55) and the 
		distribution 6 in Table \ref{Controls}. 
	\end{itemize}
	
\noindent{\bf{Couple without child}}
	\begin{itemize}
		\item The age of the individual 1 is determined using the distribution 4 in Table \ref{Controls}.
		\item The age of the individual 2 is determined using the distribution 5 in Table \ref{Controls}.
	\end{itemize}
	
\noindent{\bf{Couple with child}}
	\begin{itemize}
		\item The age of the individual 1 is determined using the distribution 4 in Table \ref{Controls}.
		\item The age of the individual 2 is determined using the distribution 5 in Table \ref{Controls}.	
		\item The ages of the children are determined according to the age of individual 1 and the distribution 6 in Table \ref{Controls}.
	\end{itemize}
	
\noindent{\bf{Other}}
	\begin{itemize}
		\item The age of the individual 1 is determined using the distribution 4 in Table \ref{Controls}.
		\item The ages of the others individuals are determined according to the age of individual 1.
	\end{itemize}
	
To obtain a synthetic population $P$ with households $Y$ filled by individuals $X$ we use the Algorithm \ref{PopGenFlo2} where we approximate the Equation \ref{eq2} with the distributions 4, 5 and 6 in Table \ref{Controls}. We put no constraint on the number of individuals in the age pyramid, hence the reference population does not give any advantage to the sample-free method. Figure S\ref{Fig3abc} and Figure S\ref{Fig4abc} show the values obtained for individual's and household's attributes for the Auvergne region and for Marsac-en-Livradois, a municipality drawn at random among the 1310 Auvergne municipalities. These figures show the results obtained with the reference, the sample-free and the sample-based populations.

\section*{Comparing sample-free and sample-based approaches}

The attributes of both individuals and households are respectivily described in Table \ref{Ind} and Table \ref{HH}. The joint-distributions of both the attributes for individuals and households give respectively the number of individuals of each individual type $n_T=\{n_{t_k}\}_{1\leq k \leq q}$ and the number of households of each household type $n_U=\{n_{u_l}\}_{1\leq l \leq p}$. In this case, $q=130$ and $p=17$. It's important to note that $p$ is not equal to $6 \cdot 5 = 30$ because we remove from the list of household types the inconsistent values like for example single households of size $5$. We do the same for the individual types (removing for example retired individuals of age comprised betweeen 0 and 5). 

\begin{table}
	\caption{Individual level attributes}
	\label{Ind}
	\begin{center}
		\begin{tabular}{|>{}m{3cm}|m{4.5cm}<{}|}
		\hline
		\textbf{Attribute} & \textbf{Value}\\
			\hline 
			Age & [0,5[\\
			 & [5,15[\\
			 & [15,25[\\
			 & [25,35[\\
			 & [35,45[\\
			 & [45,55[\\
			 & [55,65[\\
			 & [65,75[\\
			 & [75,85[\\
			 & 85 and more\\
			\hline
			Activity Status & Student\\
			 & Active\\		
			\hline
			Family Status & Head of a single household\\ 
			& Head of a monoparental household\\ 
			& Head of a couple without children household\\ 
			& Head of a couple with children household\\ 
			& Head of a other household\\ 
			& Child of a monoparental household\\ 
			& Child of a couple with children household\\ 
			& Partner\\ 
			& Other\\ 
			\hline
		\end{tabular}
	\end{center}
\end{table}

\subsection*{Fitting accuracy measures}

We need fitting accuracy measures to evaluate the adequacy between both observed $O$ and estimated $E$ household and individual distributions. The first measure is the Proportion of Good Prediction (PGP) (Equation \ref{PGP}), we choose this first indicator for the facility of interpretation. In the Equation \ref{PGP} we multiplied by 0.5 because as we have $\sum_{k=1}^p O_k =\sum_{k=1}^p E_k$, each misclassified individual or household is counted twice \cite{Harland2012}.

\begin{equation}
	PGP=1-\frac{1}{2}\frac{\sum_{k=1}^p |O_k-E_k|}{\sum_{k=1}^p O_k}
\label{PGP} 
\end{equation}

We use the $\chi^2$ distance to perform a statistic test. Obviously the modalities with a zero value for the observed distribution are not included in the $\chi^2$ computation. If we consider a distibution with $p$ modalities different from zero in the observed distribution, the $\chi^2$ distance follows a $\chi^2$ distribution with $p-1$ degrees 

\begin{table}
	\caption{Household level attributes}
	\label{HH}
	\begin{center}
		\begin{tabular}{|l|l|}
		\hline
		\textbf{Attribute} & \textbf{Value}\\
			\hline 
			Size & 1 individual\\
			& 2 individuals\\
			& 3 individuals\\
			& 4 individuals\\
			& 5 individuals\\
			& 6 and more individuals\\
			\hline
			Type & Single\\
			& Monoparental\\
			& Couple without children\\
			& Couple with children\\
			& Other\\
			\hline
		\end{tabular}
	\end{center}
\end{table}

of freedom. 

\begin{equation}
	\chi^2=\frac{\sum_{k=1}^p (O_k-E_k)^2}{\sum_{k=1}^p O_k}
\label{khi} 
\end{equation}

For more details on the fitting accuracy measures see \cite{Voas2001}.

\begin{figure*}
		\includegraphics[width=\linewidth]{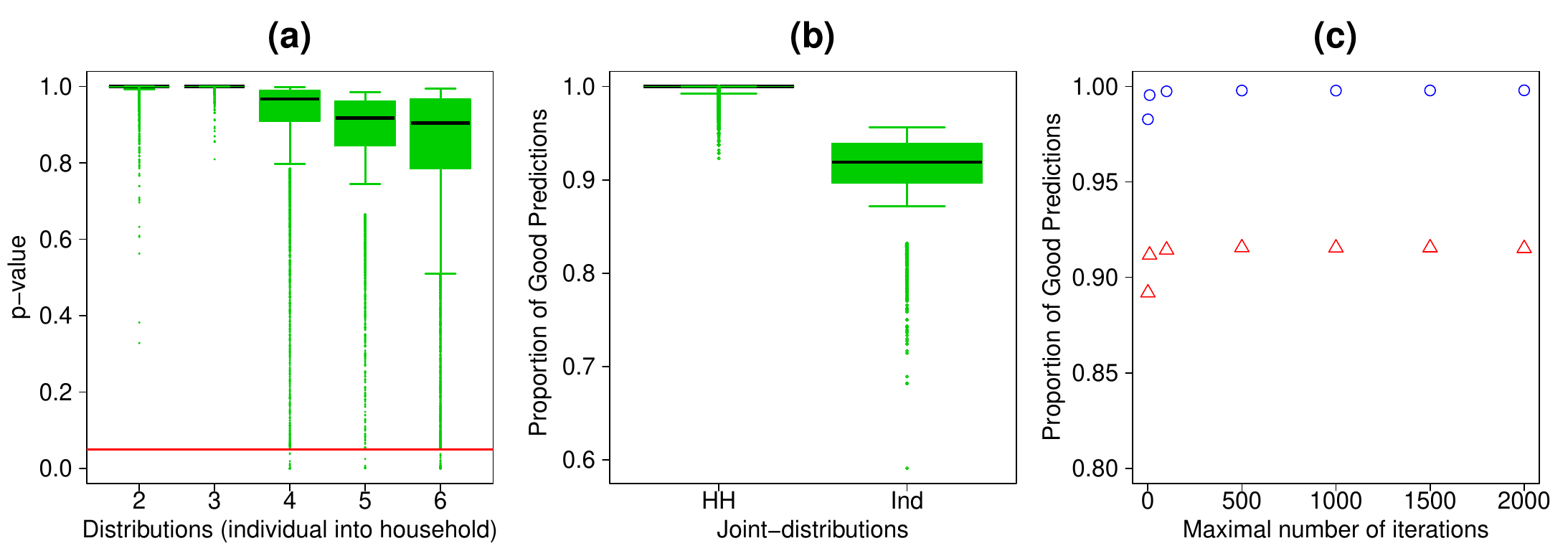}
		\caption{(a) Boxplots of the p-values obtained with the $\chi^2$ distance between the estimated distributions and the observed distributions for 
		each distribution for affecting individual into household, municipalities and replications. The x-axis represents the distributions presented in Table \ref{Controls}. The red line 
		represents the risk 5\% for the $\chi^2$ test. (b) Boxplots of the proportion of good predictions for each joint-distribution, municipalities and replications. (c) Average  
		proportion of good predictions as a function of the number of maximal iteration by households. Blue circles for the households. Red triangles for the individual.  
		\label{Fig1}}
\end{figure*}

\begin{figure*}
		\includegraphics[width=\linewidth]{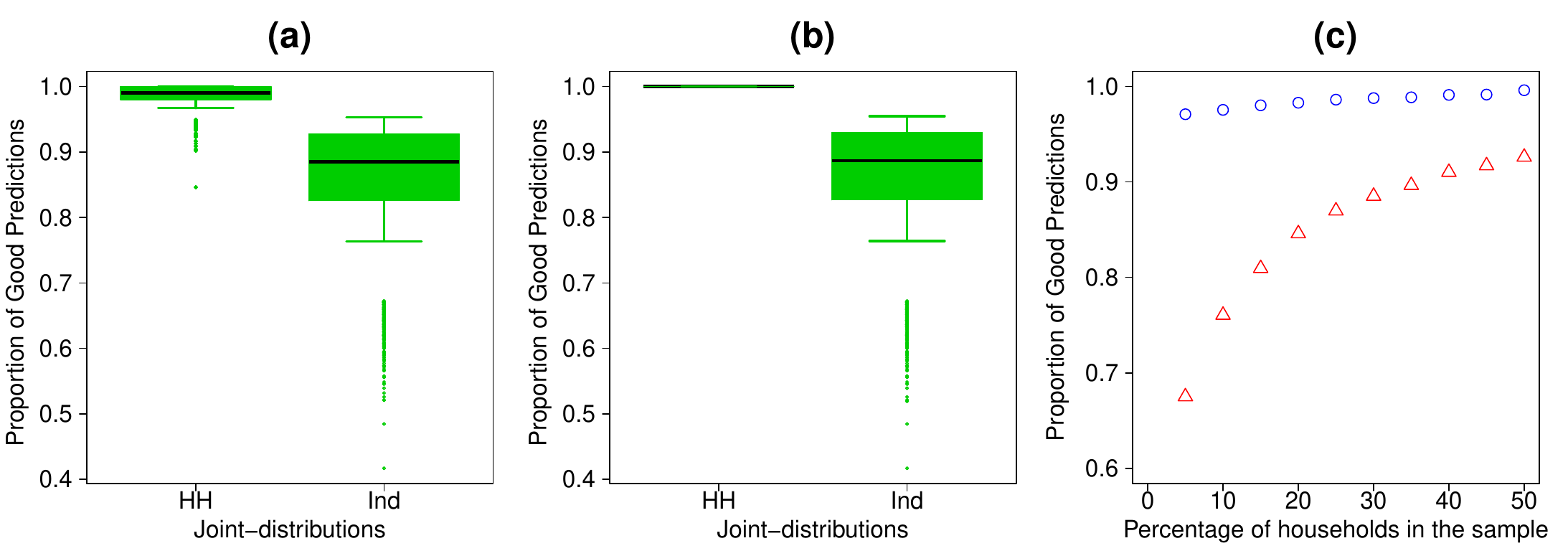}
		\caption{(a) Boxplots of the proportion of good predictions for a comparison between the estimated distribution and the observed distribution for 
		each municipality and replication. (b) Boxplots of the proportion of good predictions for a comparison between the estimated distribution and 
		the IPF-objective distribution for each municipality and replication. (c) Average proportion of good predictions  as a function of the sample size. Blue circles for the households. Red triangles for the individuals. 
		\label{Fig2}}
\end{figure*}

\subsection*{Sample-free approach}

To test the sample-free approach, we extract from the reference population, for each municipality, the distributions presented in Table \ref{Controls}. Then we use the procedure used for generating the population of reference but now with the constraints on the number of individuals from the age pyramid derived from the reference (remember that we did not have such constraints when generating the reference population). Then we fill the households with the individuals one at a time using the distributions for affecting individual into household. We limit the number of iterations to 1000 trials by household: If after 1000 trials a household is not filled, we put at random individuals in this household and we change its type to "other". We repeat the process 100 times and we choose, for each municipality, the synthetic population minimizing the $\chi^2$ distance between simulated and reference distributions for affecting individual into household. 

In order to assess the robustness of the stochastic sample-free approach, we generate 10 synthetic populations by municipalities, yielding 13,100 synthetic municipality populations in total. For each of them and for each distributions for affecting individual into household we compute the p-value associated to $\chi^2$ distance between the reference and estimated distributions. As we can see in the Figure \ref{Fig1} a the algorithm is quite robust.
 
To validate the algorithm we compute the proportion of good predictions for each 13,100 synthetic populations and for each joint-distribution. We obtain an average of 99.7\% of good predictions for the household distribution and 91.5\% of good predictions for the individual distribution (Figure \ref{Fig1}b). We also compute the p-value of the $\chi^2$ distance between the estimated and reference distributions for each of the synthetic populations and for each joint-distribution. Among the 13,100 synthetic populations 100\% are statistically similar to the observed one at a 0.95\% level of confidence for the household joint-distribution and 94\% for the individual joint-distribution.

In order to understand the effect of the maximal number of iterations by household, we repeat the previous tests for different values of this parameter (1,10,100,500,1000,1500 and 2000)and we compute the mean proportion of good predictions obtained for both individual and household. We note that after 100 the quality of the results no longer changes.

\begin{figure*}
  \includegraphics[scale=0.25]{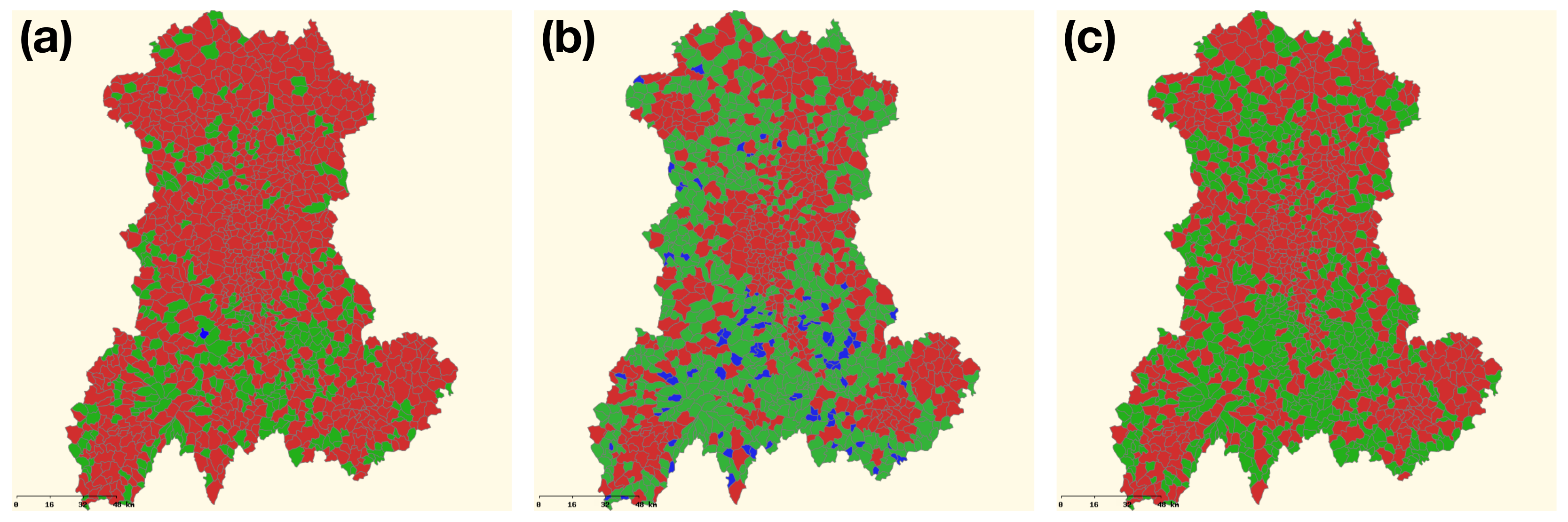}        
  \caption{Maps of the average proportion of good predictions ((a) sample-free and (b) IPU) and the number of inhabitants ((c)) 
  by municipality for the Auvergne case study. For (a)-(b), in blue $0.5\leq$ PGP $<0.75$; In green $0.75\leq$ PGP $<0.9$; In red $0.9\leq$ PGP. For (c), in green, the number of 
  inhabitants is lower than 350. In red, the number of inhabitants is upper than 350. \textit{\scriptsize{Base maps source: Cemagref - DTM - 
  D\'{e}veloppement Informatique Syst\`{e}me d'Information et Base de Donn\'{e}es : F.Bray \& A.Torre
  IGN (G\'{e}ofla\textsuperscript{{\fontsize{5}{5}\textregistered}}, 2007).}}}
  \label{Fig3}
\end{figure*}

\subsection*{IPU}

To use the IPU algorithm we need a sample of filled households and marginal variables. In order to obtain these data we pick at random a significant sample of $25\%$ of households from the reference population $P$ and we also extract from $P$ the two one-dimensional marginals (Size and Type distributions) that we need to build the household joint-distributions with IPF and the three two-dimensional marginals (Age x Activity Status, Age x Family Status and Family Status x Activity Status) joint-distributions that we need to build the individual joint-distributions with IPF. Then we apply the Algorithm \ref{IPU} using the recommendation of \cite{Ye2009} for the well-know zero-cell and zero-marginal problems to obtain a weighted sample $P_s$. With this sample we generate 100 times the synthetic population $P$ and choose the one with lowest $\chi^2$ distance between reference and simulated individual joint-distributions.

To check the results obtained with the IPU approach, we generate 10 synthetic populations by municipality using different samples of $25\%$ of households randomly selected. For each of these synthetic populations and for each joint-distribution we compute the proportion of good predictions (Figure \ref{Fig2}a). We obtain an average of 98.6\% of good predictions for the household distribution and 86.9\% of good predictions for the individual distribution. To determine the error of estimation due to the IPF procedure we compute the proportion of good predictions for the estimated and the IPF-reference distributions. As we can see in Figure \ref{Fig2}b the results are improved for the household distribution but not for the individual distribution. We also compute the p-value of the $\chi^2$ distance between the estimated and observed distributions for each of the synthetic populations and for each joint-distribution. Among the 13,100 synthetic populations 100\% are statistically similar to the observed one at a 0.95\% level of confidence for the household joint-distribution and 61\% for the individual joint-distribution. We obtained a similarity between the estimated and the IPF-objective distributions of 100\% at a 0.95\% level of confidence for the household distribution and 64\% for the individual distribution.

In order to check the sensitivity of the results to the size of the sample, we plot, on Figure \ref{Fig2}c, the average proportion of good predictions of the 13,100  household and individuals joint-distributons for different values of the percentage of the reference households drawn at random in the sample (5, 10, 15, 20 ,25, 30, 35, 40, 45 and 50). We note that the results are always good for the household distribution but for the individuals the results are good only from random sample of at least 25\% of the reference household population. Not surprisingly, globally the quality of the results increases with the parameter.

\section*{Discussion}

The sample-free method is less data demanding but it requires more data pre-processing. Indeed, this approach requires to extract the distributions for affecting individual into household from data. The sample-free method gives better fit between observed and simulated distribution for both household and individual distribution than the IPU approach. We can observe in Figure \ref{Fig3} that, for both methods, the goodness-of-fit is negatively correlated with the number of inhabitants. This observation is especially true for the IPU method because it depends on the number of individuals in the sample. Indeed, the lower is the number of individuals, the higher is the number of sparse cells in the individual distribution. The results obtained with the IPU approach depend of the quality of the initial sample. The execution time on a desktop machine (PC Intel 2.83 GHz) is almost the same for 100 maximal iterations by household for the sample-free method and 25\% reference households drawn at random in the sample reference households for the sample-based approach. 

To conclude, the sample-free method gives globally better results in this application on small French municipalities. These results confirm those of \cite{Barthelemy2012} who compared their sample-free method for working with data from different sources with a sample-based method \cite{Guo2007}, and obtained similar conclusions. Of course, these conclusions cannot be generalized to all sample-free and sample-based methods without further investigation. However, these results confirm the possibility to initialise accurately micro-simulation (or agent-based) models, using widely available data (and without any sample of households).

\begin{table}[!ht]
	\caption{Average execution time for the two approaches for different parameter values.}
	\label{Time}
		\begin{center}
			\begin{tabular}{|c|c|c|c|}
	  		\hline
	  		\multicolumn{2}{|c|}{\textbf{IPU}} &  \multicolumn{2}{c|}{\textbf{Iterative}}\\ 
				\hline
				\textbf{Sample size} & \textbf{Time} & \textbf{Iterations} & \textbf{Time}\\
				\hline
				5	  &	13min	&	1	    &	40min	\\
				10	&	24min	&	10	  &	41min	\\
				15	&	29min	&	100	  &	45min	\\
				20	&	38min	&	500	  &	58min	\\
				25	&	45min	&	1000	&	66min \\
				30	&	53min	&	1500	&	78min \\
				40	&	74min	&	2000	&	88min	\\
				\hline
	  	\end{tabular}
	  \end{center}
\end{table}

\section*{Acknowledgements}

This publication has been funded by the Prototypical policy impacts on multifunctional activities in rural municipalities collaborative project, European Union 7th Framework Programme (ENV 2007-1), contract no. 212345. The work of the first author has been funded by the Auvergne region.

\bibliographystyle{unsrt}  
\bibliography{PopGen}

\begin{thebibliography}{10}

\bibitem{Gargiulo2010}
F.~Gargiulo, S.~Ternes, S.~Huet, and G.~Deffuant.
\newblock An iterative approach for generating statistically realistic
  populations of households.
\newblock {\em PLoS ONE}, 5, 2010.

\bibitem{Ye2009}
X.~Ye, K.~Konduri, R.~M. Pendyala, B.~Sana, and P.~Waddell.
\newblock A methodology to match distributions of both household and person
  attributes in the generation of synthetic populations.
\newblock In {\em 88th Annual Meeting of the Transportation Research Board},
  2009.

\bibitem{Wilson1976}
A.~G. Wilson and C.~E. Pownall.
\newblock A new representation of the urban system for modelling and for the
  study of micro-level interdependence.
\newblock {\em Area}, 8(4):246--254, 1976.

\bibitem{Deming1940}
W.~E. Deming and F.~F. Stephan.
\newblock On a least squares adjustment of a sample frequency table when the
  expected marginal totals are known.
\newblock {\em Annals of Mathematical Statistics}, 11:427--444, 1940.

\bibitem{Beckman1996}
R.~J. Beckman, K.~A. Baggerly, and M.~D. McKay.
\newblock Creating synthetic baseline populations.
\newblock {\em Transportation Research Part A: Policy and Practice}, 30(6 PART
  A):415--429, 1996.

\bibitem{Huang2002}
Z.~Huang and P.~Williamson.
\newblock A comparison of synthetic reconstruction and combinatorial
  optimization approaches to the creation of small-area microdata.
\newblock Working paper, Departement of Geography, University of Liverpool,
  2002.

\bibitem{Guo2007}
J.~Y. Guo and C.~R. Bhat.
\newblock Population synthesis for microsimulating travel behavior.
\newblock {\em Transportation Research Record: Journal of the Transportation
  Research Board}, 2014:92--101, 2007.

\bibitem{Arentze2007}
T~Arentze, H~Timmermans, and F~Hofman.
\newblock Creating synthetic household populations: Problems and approach.
\newblock {\em Transportation Research Record: Journal of the Transportation
  Research Board}, 2014:85--91, 2007.

\bibitem{Voas2000}
D.~Voas and P.~Williamson.
\newblock An evaluation of the combinatorial optimisation approach to the
  creation of synthetic microdata.
\newblock {\em International Journal of Population Geography}, 6(5):349--366,
  2000.

\bibitem{Barthelemy2012}
P.~Barthelemy, J.and~Toint.
\newblock Synthetic population generation without a sample.
\newblock {\em Transportation Science}, 47:266--279, 2013.

\bibitem{Harland2012}
K.~Harland, A.~Heppenstall, D.~Smith, and M.~Birkin.
\newblock Creating realistic synthetic populations at varying spatial scales: A
  comparative critique of population synthesis techniques.
\newblock {\em Journal of Artificial Societies and Social Simulation}, 15(1):1,
  2012.

\bibitem{Voas2001}
D.~Voas and P.~Williamson.
\newblock Evaluating goodness-of-fit measures for synthetic microdata.
\newblock {\em Geographical and Environmental Modelling}, 5(2):177--200, 2001.

\end{thebibliography}

\makeatletter
\renewcommand{\fnum@figure}{\small\textbf{\figurename~S\thefigure}}
\makeatother
\setcounter{figure}{0}

\onecolumngrid

\begin{figure}[!ht]
  \centering
  \includegraphics[scale=0.25]{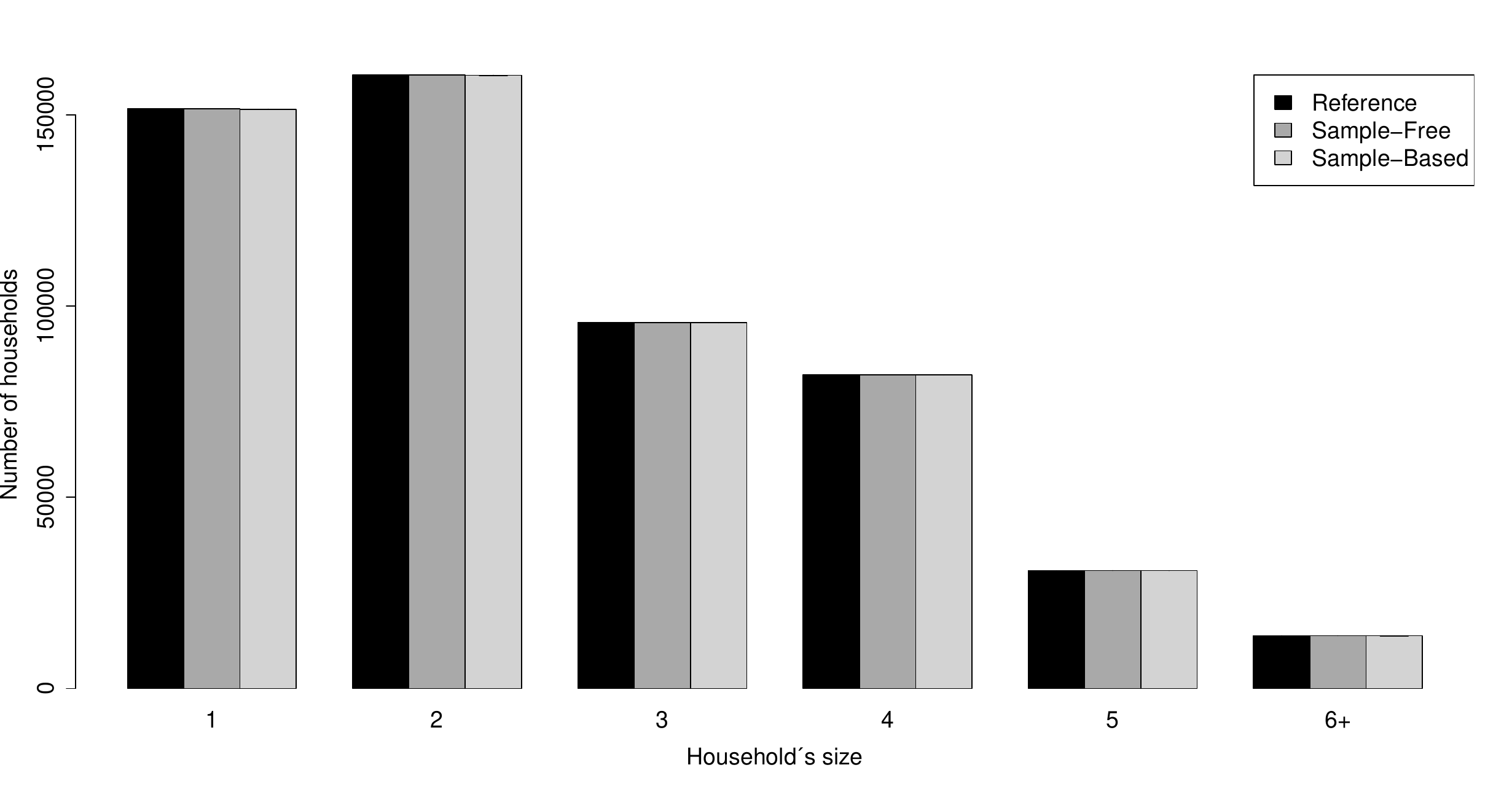}\\$$(a)$$             
  \includegraphics[scale=0.25]{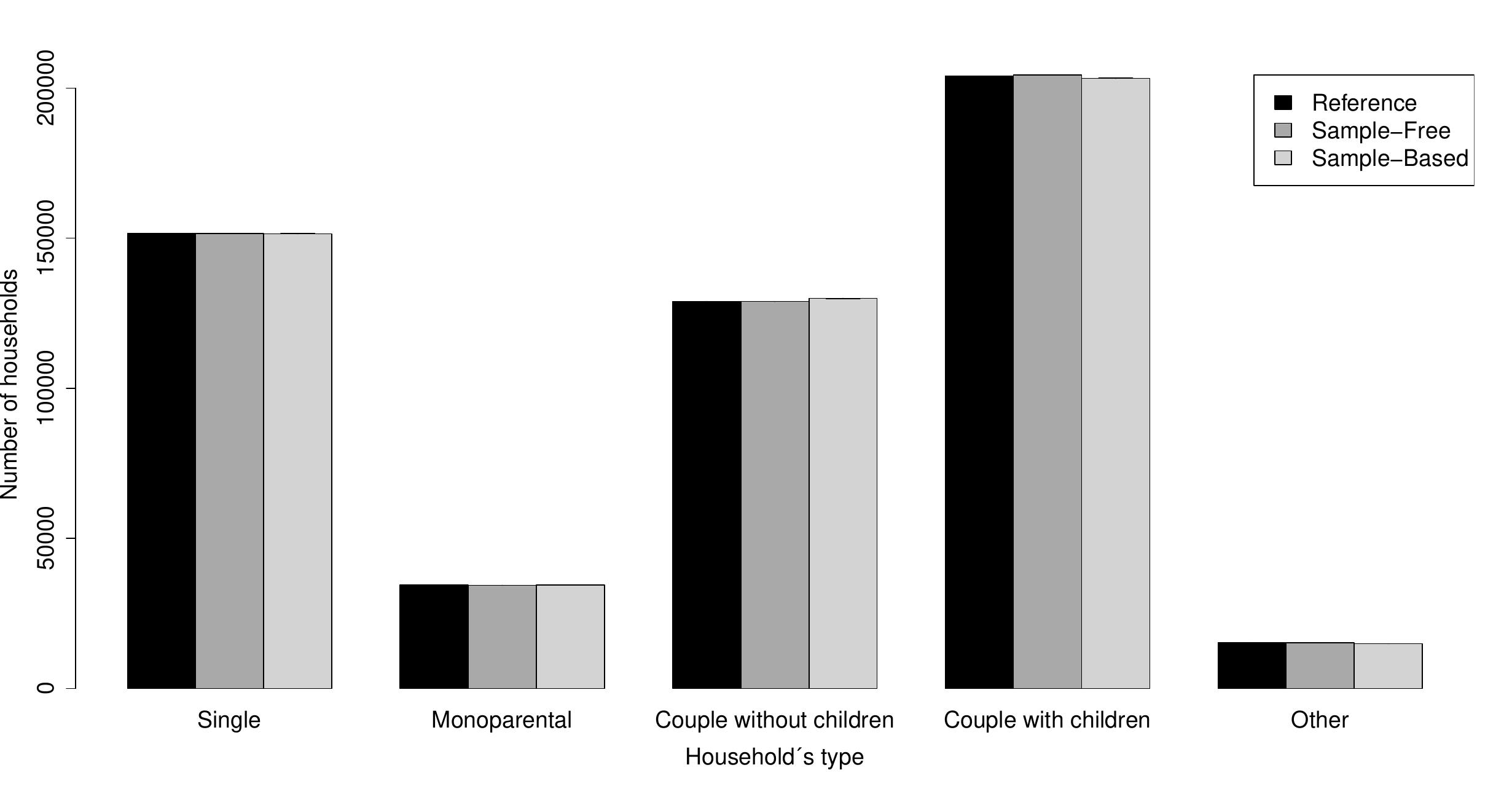}\\$$(b)$$
  \includegraphics[scale=0.25]{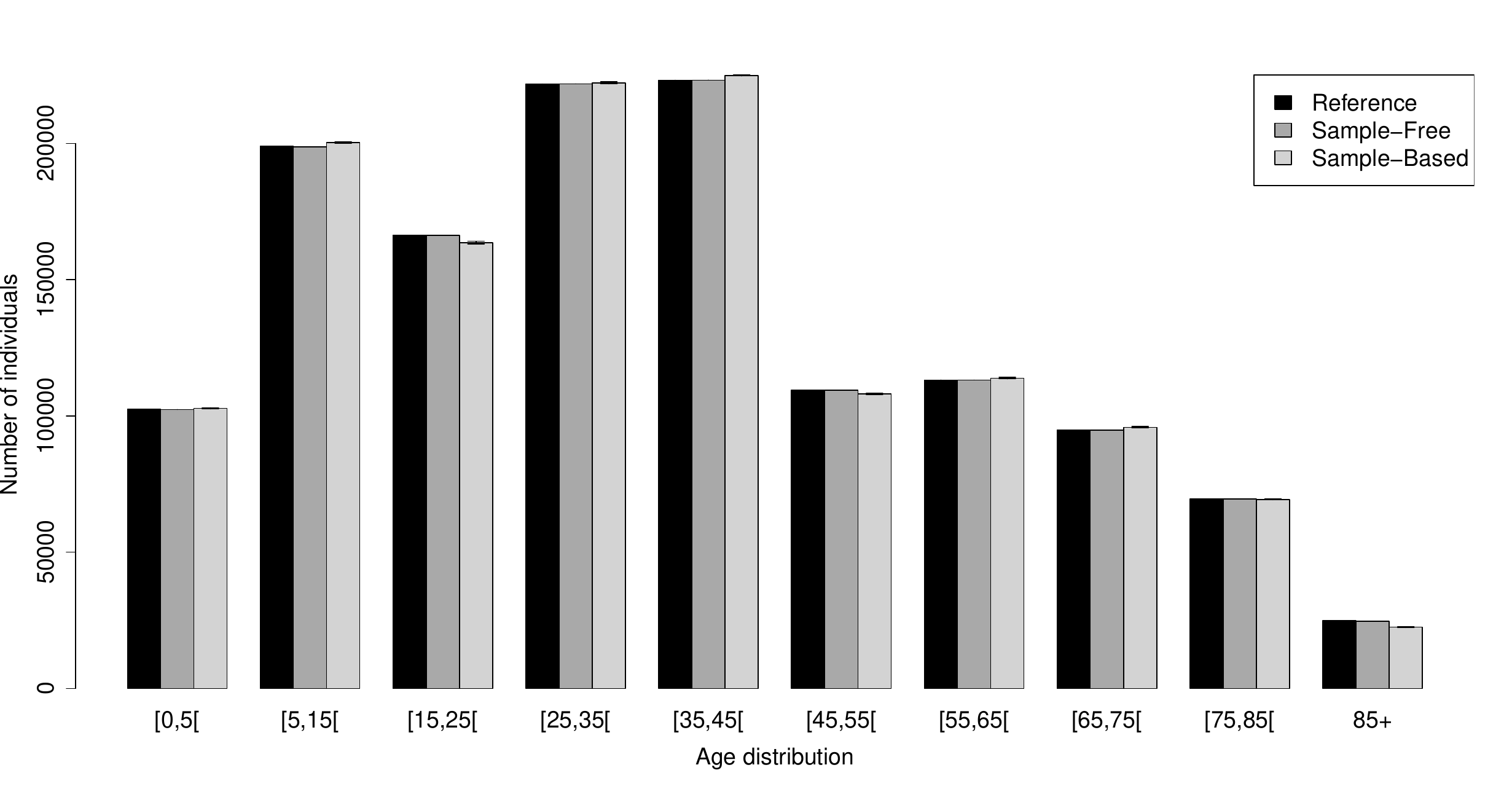}\\$$(c)$$
   \caption{Barplots of individual's and household's attributes for the Auvergne region. (a) Household's size. (b) Household's type. (c) Individual's age distribution. In black, the reference population. In dark grey, the population obtained with the sample-free method (1000 maximal iterations). In light grey, the population obtained with the sample-based method (25\% of the reference household population). The bars represent the standard deviations obtained with 10 replications.}
  \label{Fig3abc}
\end{figure}

\begin{figure}[!ht]
  \centering
  \includegraphics[scale=0.25]{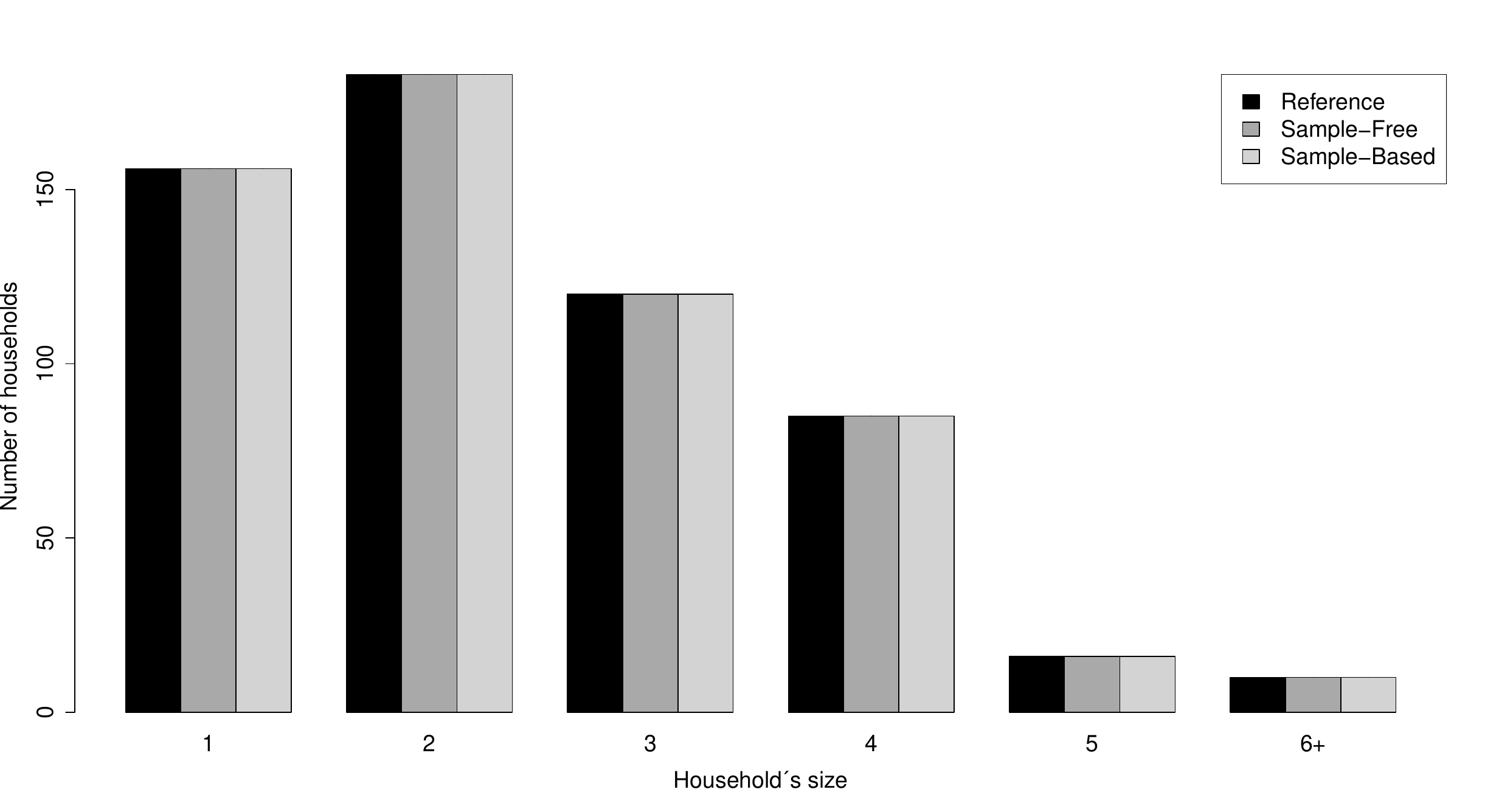}\\$$(a)$$             
  \includegraphics[scale=0.25]{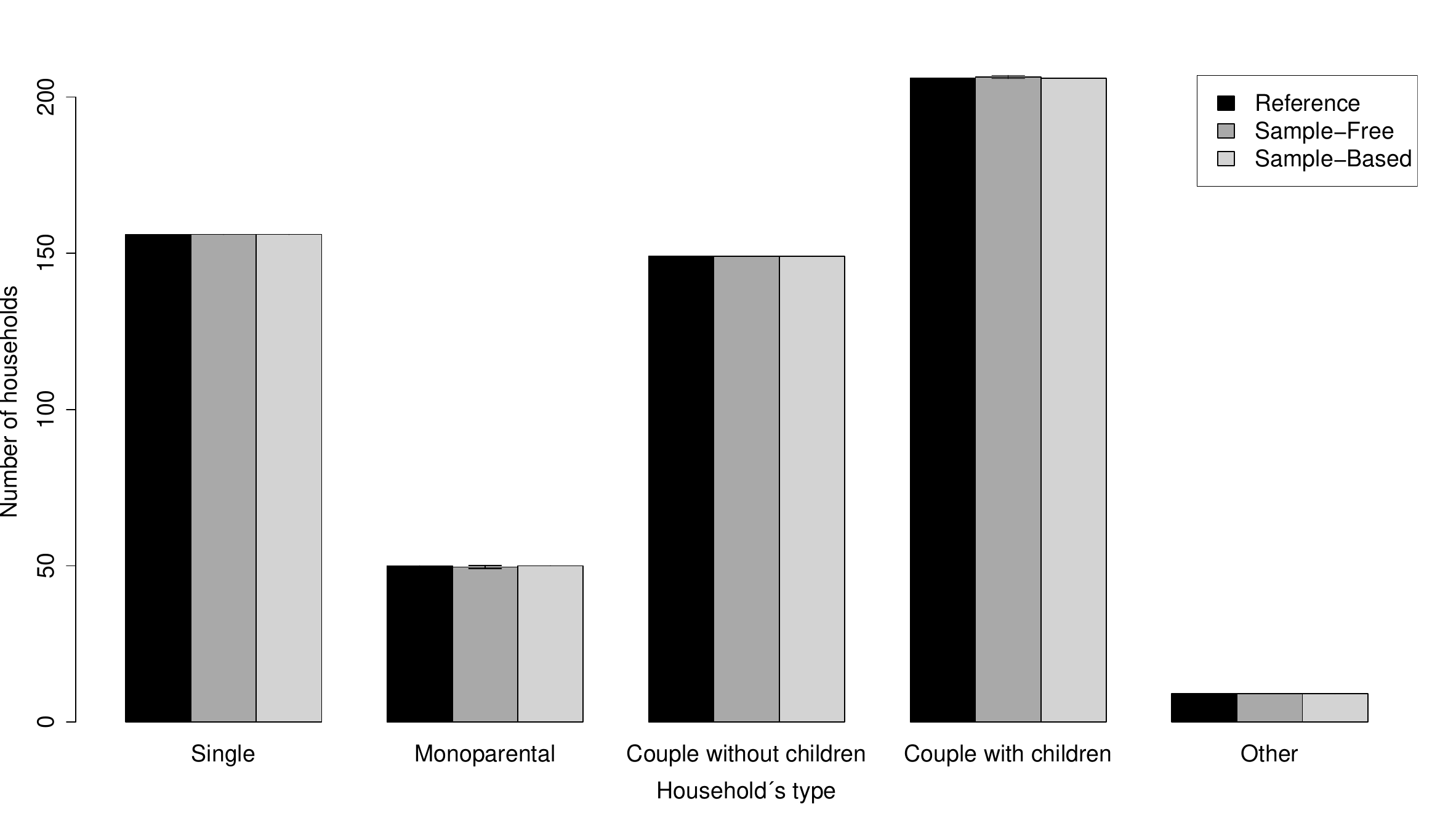}\\$$(b)$$
  \includegraphics[scale=0.25]{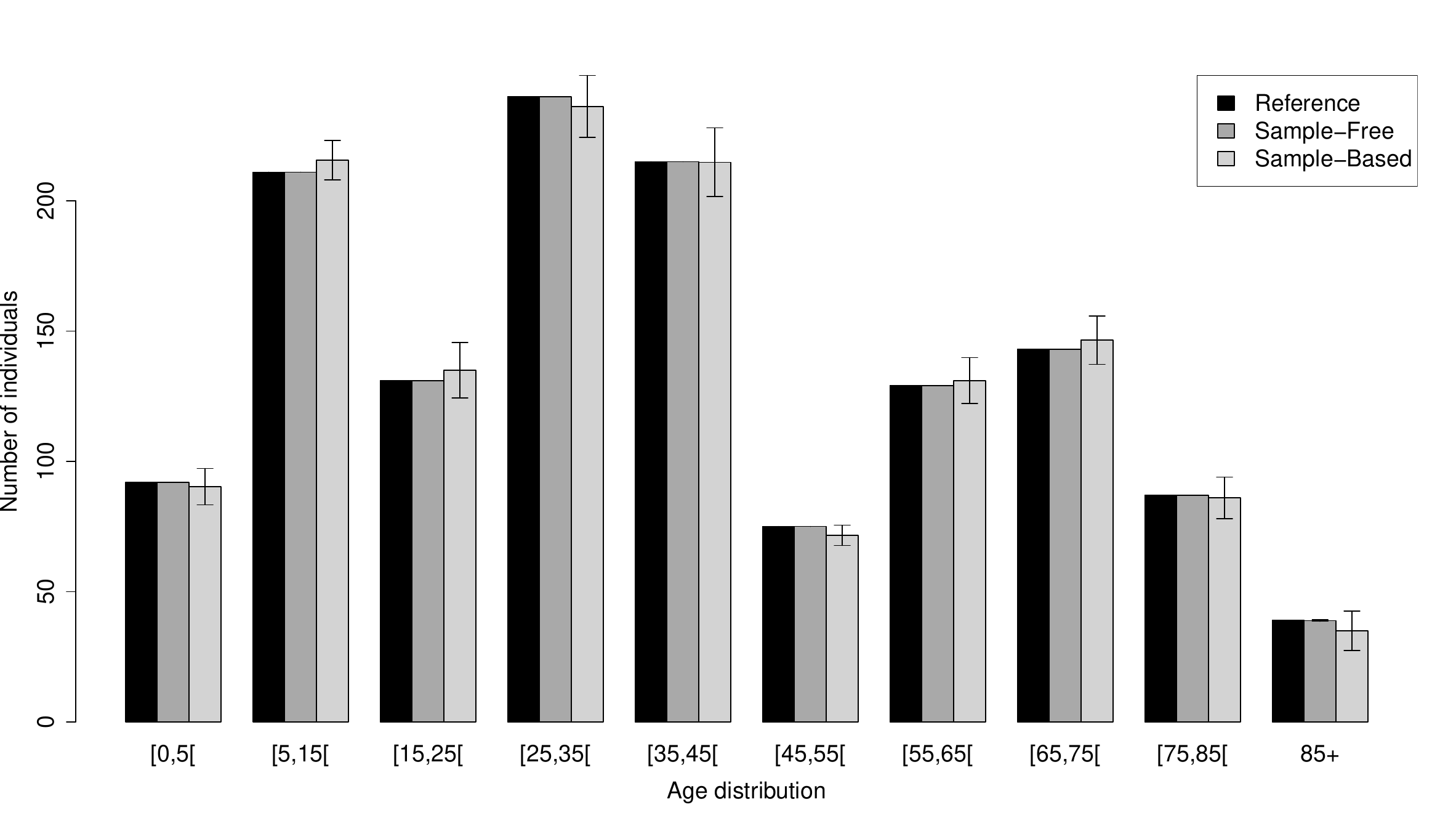}\\$$(c)$$
   \caption{Barplots of individual's and household's attributes for Marsac-en-Livradois, a municipality drawn at random among the 1310 Auvergne municipalities. (a) Household's size. (b) Household's type. (c) Individual's age distribution. In black, the reference population. In dark grey, the population obtained with the sample-free method (1000 maximal iterations). In light grey, the population obtained with the sample-based method (25\% of the reference household population). The bars represent the standard deviations obtained with 10 replications.}
  \label{Fig4abc}
\end{figure}

\end{document}